\pdfoutput=1
\documentclass[aps,prd,amsmath,floats,floatfix, twocolumn,
superscriptaddress,nofootinbib,showpacs]{revtex4-1}

\usepackage[T1]{fontenc}
\usepackage[utf8]{inputenc}
\usepackage{lmodern}
\usepackage[caption=false]{subfig}
\usepackage{verbatim}

\usepackage[dvipsnames, usenames]{xcolor}
\definecolor{linkcolor}{rgb}{0.0,0.3,0.5}
\usepackage[hypertexnames=false, unicode, colorlinks=true, linkcolor=linkcolor,
citecolor=linkcolor, filecolor=linkcolor,urlcolor=linkcolor,
pdfusetitle]{hyperref}

\usepackage[all]{hypcap}
\usepackage{graphicx}
\usepackage{tikz}
\usepackage{xspace}
\usepackage{amssymb}
\usepackage[normalem]{ulem} 
\usepackage{bm} 
\usepackage{enumitem,amssymb}

\usepackage{microtype}
\usepackage{orcidlink}

\usepackage[english]{babel}
\usepackage{blindtext}

\graphicspath{%
  {figs/}%
}

\DeclareMathAlphabet{\mathpzc}{OT1}{pzc}{m}{it}

\newcommand{\etal}{\textit{et al.\ }}

\newlist{todolist}{itemize}{2}
\setlist[todolist]{label=$\square$}

\newcommand{\Chandra}{\texttt{Chandra}~}

\begin{document}

\title{Foundations of Direct Waves in Schwarzschild Ringdown}


\newcommand\Caltech{\affiliation{TAPIR 350-17, California Institute of
    Technology, 1200 E California Boulevard, Pasadena, CA 91125, USA}}
\newcommand{\Perimeter}{\affiliation{Perimeter Institute for Theoretical Physics, Waterloo, ON N2L2Y5, Canada}}

\author{Sizheng Ma\orcidlink{0000-0002-4645-453X}}
\email{sma2@perimeterinstitute.ca}
\Perimeter

\author{Hai-Yang Wang\orcidlink{0000-0001-7167-6110}}
\affiliation{California Institute of Technology, Pasadena, CA 91125, USA}

\hypersetup{pdfauthor={Ma et al.}}

\date{\today}

\begin{abstract}
Recent studies have identified a new component in black-hole ringdown from merging binaries, termed the \emph{direct wave}. This component was argued to be tied to the dynamical source evolution near the black-hole horizon, and thus to encode horizon information. Yet a firm theoretical foundation for the direct wave has been lacking. Here we fill this gap by deriving direct waves from first principles in Schwarzschild spacetime, using the causal structure of the Green's function. We show that the direct wave does not vanish and is governed by the near-horizon source dynamics. Our results establish a theoretical basis for direct waves as a probe of near-horizon dynamics, complementary to quasinormal modes.
\end{abstract}

\maketitle

\section{Introduction} 
\label{sec:introduction}
Binary-black-hole coalescence is commonly divided into inspiral, merger, and ringdown. Understanding each stage is essential for extracting physics from gravitational waves (GWs) and for testing general relativity in the strong-field regime~\cite{TheLIGOScientific:2016src,LIGOScientific:2019fpa,LIGOScientific:2021sio,LIGOScientific:2026qni}. The ringdown, in particular, has long been modeled by quasinormal modes (QNMs)~\cite{Berti:2009kk,Berti:2025hly}, the characteristic damped oscillations of the remnant black hole (BH), whose measurement underpins BH spectroscopy~\cite{Dreyer:2003bv,Buonanno:2006ui,Berti:2005ys,Berti:2018vdi,Giesler:2019uxc,Giesler:2024hcr,Isi:2019aib,Carullo:2019flw,Cotesta:2022pci,Isi:2022mhy,Finch:2022ynt,Ma:2023vvr,Capano:2021etf,Siegel:2023lxl,LIGOScientific:2025rsn,Wang:2025rvn,LIGOScientific:2025slb,LIGOScientific:2025wao,LIGOScientific:2026wpt,Dyer:2025hdt,Dyer:2025iwj}.

In recent years, sustained efforts have pushed beyond this description, identifying additional structures in merger--ringdown waveforms: nonlinear effects~\cite{Mitman:2022qdl,Cheung:2022rbm,Ma:2022wpv,Khera:2024bjs,Ma:2024qcv,Zhu:2024rej,Redondo-Yuste:2023seq,Bucciotti:2024zyp,Bourg:2024jme,Dyer:2025hdt}, dynamical QNM excitation~\cite{Andersson:1996cm,Chavda:2024awq,DeAmicis:2025xuh}, prompt response~\cite{Ma:2026qbq}, and tails~\cite{DeAmicis:2024eoy,Islam:2024vro,DeAmicis:2024not,Ma:2024hzq,Cardoso:2024jme,Islam:2025wci}. Notably, a \emph{direct-wave} component was found in numerical-relativity (NR) waveforms~\cite{Oshita:2025qmn,Lu:2025vol}: once the dominant QNMs are removed with rational filters~\cite{Ma:2022wpv,Ma:2023vvr,Ma:2023cwe,Lu:2025mwp}, several oscillatory, decaying cycles remain near the strain peak, a finding since confirmed independently~\cite{Kankani:2026byb,Chung:2026eph,Dyer:2026yex}.

Inspired by earlier studies of horizon modes~\cite{Mino:2008at,Zimmerman:2011dx}, Oshita \etal \cite{Oshita:2025qmn} interpreted the direct wave as arising from source motion near the horizon: its frequency tracks the source's orbital motion, modulated by the horizon angular frequency $\Omega_H$, while its decay rate is set by the radial infall velocity, imprinted with the surface gravity $\kappa$. Motivated by this picture, Lu \etal \cite{Lu:2025vol} identified this component in the GW event GW250114~\cite{LIGOScientific:2025slb}, with frequency and decay rate broadly consistent with NR expectations --- close to, though not exactly at, $\Omega_H$ and $\kappa$. If correct, this interpretation makes direct waves an observational route to near-horizon dynamics during and after BH mergers.

Yet the theoretical foundation of direct waves remains unsettled: the interpretation of Oshita \etal \cite{Oshita:2025qmn} rests on a simplified semianalytical model, and a first-principles derivation is lacking. Kuntz \etal \cite{Kuntz:2026xep} recently sharpened the issue by arguing that horizon modes vanish in Schwarzschild spacetime and extrapolating this conclusion to direct waves. It is therefore timely to place direct waves on a firm theoretical footing.

Here we do so by analyzing the causal structure of rationally filtered waveforms, building on the recently developed decomposition of the Schwarzschild Green's function~\cite{Arnaudo:2025uos,Su:2026fvj,Kuntz:2025gdq,Rosato:2026moe,Ma:2026prompt}. The method decomposes filtered waveforms into physically distinct components and identifies the origin of the direct waves of \cite{Oshita:2025qmn,Lu:2025vol}. As a proof of principle, we study the Zerilli equation sourced by plunging point particles.

\section{Causal structure of the filtered Green's function}
\label{sec:filtered_green}
We consider the sourced Zerilli equation
\begin{equation}
\left[
-\partial_t^2
+\partial_{r_*}^2
-V_\ell^{\mathrm Z}(r)
\right]
\Psi_{\ell m}(t,r)
=
S_{\ell m}(t,r),
\end{equation}
where $ r_*=r+2\ln\left(\frac{r}{2}-1\right)$; throughout, we use geometric units $G=c=1$ and set the BH mass to unity.
At future null infinity, the waveform can be written as a convolution of the source with the retarded Green's function $G_{\ell}$,
\begin{align}
    \Psi_{\ell m}(u)=\int dt\, \int dr\,  G_{\ell}\bigl(u - t; r\bigr) S_{\ell m}(t,r), \label{eq:Psi_Zerilli_convolution}
\end{align}
with $u$ the retarded time. The frequency-domain representation of $G_{\ell}$ is
\begin{align}
  G_\ell(u;r) = \int_{-\infty}^{\infty}\frac{d\omega}{2\pi}\,
  e^{-i\omega u}\,\frac{R^{\rm in}_\omega(r)}{2i\omega A^{\rm inc}_{\rm in}}.
  \label{eq:G_ell_def}
\end{align}
Here $R^{\rm in}_{\omega}$ is the homogeneous in-mode solution, obeying
\begin{align}
  R^{\rm in}_\omega(r) &\sim
  \begin{cases}
    e^{-i\omega r_*}, & r_*\to-\infty,\\
    A^{\rm ref}_{\rm in}e^{i\omega r_*}+A^{\rm inc}_{\rm in}e^{-i\omega r_*}, & r_*\to+\infty.
  \end{cases}
  \label{eq:in_solution}
\end{align}

To remove a prescribed set of QNMs $\{\omega_n\}$, we apply the rational filter \cite{Ma:2022wpv,Ma:2023cwe,Ma:2023vvr}
\begin{align}
  F(\omega) = \prod_{n=0}^N
  \frac{\omega-\omega_n}{\omega-\bar\omega_n}\,
  \frac{\omega+\bar\omega_n}{\omega+\omega_n},
  \label{eq:rational_filter}
\end{align}
where $\bar\omega_n$ denotes the complex conjugate.
The filtered Green's function is then 
\begin{align} 
G^{\rm filt}_\ell(u;r) = \int_{-\infty}^{\infty} \frac{d\omega}{2\pi}\, e^{-i\omega u}\, \frac{ F(\omega) R^{\rm in}_{\omega}(r)} {2i\omega A^{\rm inc}_{\rm in}} . \label{eq:G_filt_def} 
\end{align} 
The corresponding filtered Zerilli waveform $\Psi^{\rm filt}_{\ell m}$ is obtained by replacing $G_\ell$ in Eq.~\eqref{eq:Psi_Zerilli_convolution} with $G^{\rm filt}_\ell$.

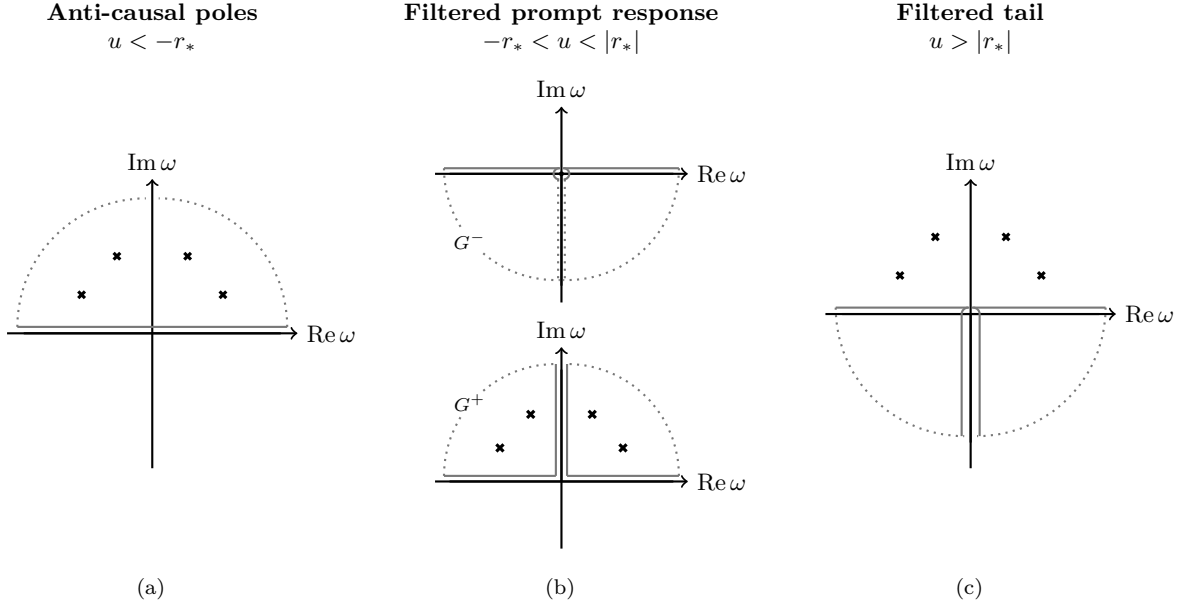
\begin{figure*}[t]
\centering
\definecolor{mygray}{RGB}{120,120,120}
\begin{tabular}{@{}c@{\hspace{0.032\textwidth}}c@{\hspace{0.032\textwidth}}c@{}}
\makebox[0.27\textwidth][c]{\textbf{Anti-causal poles}} &
\makebox[0.27\textwidth][c]{\textbf{Filtered prompt response}} &
\makebox[0.27\textwidth][c]{\textbf{Filtered tail}} \\
\makebox[0.27\textwidth][c]{$u<-r_*$} &
\makebox[0.27\textwidth][c]{$-r_*<u<|r_*|$} &
\makebox[0.27\textwidth][c]{$u>|r_*|$} \\[0.5em]
\subfloat[\label{fig:uhp_no_pia}]{\raisebox{1.20cm}{\makebox[0.27\textwidth][c]{
\begin{tikzpicture}[scale=1.70]
  \path[use as bounding box] (-1.25,-1.10) rectangle (1.25,1.25);
  \draw[->, thick] (-1.13,0) -- (1.13,0) node[right] {\small $\mathrm{Re}\,\omega$};
  \draw[->, thick] (0,-1.05) -- (0,1.2) node[above] {\small $\mathrm{Im}\,\omega$};
  \draw[thick] (-1,0) -- (1,0);
  \definecolor{mygray}{RGB}{120,120,120}
  \draw[dotted, thick, mygray] (-1.05,0.05) arc[start angle=180, end angle=90, radius=1];
  \draw[dotted, thick, mygray] (1.05,0.05) arc[start angle=0, end angle=90, radius=1];
  \draw[thick, mygray] (-1.05,0.05) -- (1.05,0.05);
  \foreach \n in {1,2}{
    \begin{scope}[shift={(0.55/\n,0.30*\n)}]
      \draw[very thick] (-0.026,-0.026) -- (0.026,0.026);
      \draw[very thick] (-0.026,0.026) -- (0.026,-0.026);
    \end{scope}
    \begin{scope}[shift={(-0.55/\n,0.30*\n)}]
      \draw[very thick] (-0.026,-0.026) -- (0.026,0.026);
      \draw[very thick] (-0.026,0.026) -- (0.026,-0.026);
    \end{scope}
  }
\end{tikzpicture}
}}}
&
\subfloat[\label{fig:filtered_contours}]{\makebox[0.27\textwidth][c]{
\begin{tikzpicture}[scale=1.48]
  \path[use as bounding box] (-1.25,-3.50) rectangle (1.25,0.70);
  \begin{scope}
    \draw[->, thick] (-1.13,0) -- (1.13,0) node[right] {\small $\mathrm{Re}\,\omega$};
    \draw[->, thick] (0,-1.15) -- (0,0.6) node[above] {\small $\mathrm{Im}\,\omega$};
    \draw[thick] (0,0) -- (0,-1);
    \draw[thick] (-1,0) -- (1,0);
    \definecolor{mygray}{RGB}{120,120,120}
    \draw[dotted, thick, mygray] (-1.05,0.05) arc[start angle=180, end angle=272, radius=1];
    \draw[dotted, thick, mygray] (1.05,0.05) arc[start angle=0, end angle=-92, radius=1];
    \draw[dotted, thick, mygray] (0.03,-0.05) -- (0.03,-0.95);
    \draw[dotted, thick, mygray] (-0.03,-0.05) -- (-0.03,-0.95);
    \draw[thick, mygray] (-1.05,0.05) -- (-0.01,0.05);
    \draw[thick, mygray] (0.01,0.05) -- (1.05,0.05);
    \draw[thick, mygray] (0.02,-0.05) arc[start angle=-90, end angle=90, radius=0.05];
    \draw[thick, mygray] (-0.02,-0.05) arc[start angle=-90, end angle=-270, radius=0.05];
    \fill (0,0) circle (0.022);
    \node[fill=white, inner sep=0.7pt] at (-0.82,-0.60) {\scriptsize $G^-$};
  \end{scope}
  \begin{scope}[yshift=-2.75cm]
    \draw[->, thick] (-1.13,0) -- (1.13,0) node[right] {\small $\mathrm{Re}\,\omega$};
    \draw[->, thick] (0,-0.6) -- (0,1.2) node[above] {\small $\mathrm{Im}\,\omega$};
    \draw[thick] (0,0) -- (0,1);
    \draw[thick] (-1,0) -- (1,0);
    \draw[dotted, thick, mygray] (-1.05,0.05) arc[start angle=180, end angle=90, radius=1];
    \draw[dotted, thick, mygray] (1.05,0.05) arc[start angle=0, end angle=90, radius=1];
    \draw[thick, mygray] (0.05,0.05) -- (0.05,1.05);
    \draw[thick, mygray] (-0.05,0.05) -- (-0.05,1.05);
    \draw[thick, mygray] (-1.05,0.05) -- (-0.05,0.05);
    \draw[thick, mygray] (0.05,0.05) -- (1.05,0.05);
    \foreach \n in {1,2}{
      \begin{scope}[shift={(0.55/\n,0.30*\n)}]
        \draw[very thick] (-0.032,-0.032)--(0.032,0.032);
        \draw[very thick] (-0.032,0.032)--(0.032,-0.032);
      \end{scope}
      \begin{scope}[shift={(-0.55/\n,0.30*\n)}]
        \draw[very thick] (-0.032,-0.032)--(0.032,0.032);
        \draw[very thick] (-0.032,0.032)--(0.032,-0.032);
      \end{scope}
    }
    \node[fill=white, inner sep=0.7pt] at (-0.82,0.70) {\scriptsize $G^+$};
  \end{scope}
\end{tikzpicture}
}}
&
\subfloat[\label{fig:unfiltered_contour}]{\raisebox{1.20cm}{\makebox[0.27\textwidth][c]{
\begin{tikzpicture}[scale=1.70]
  \path[use as bounding box] (-1.25,-1.25) rectangle (1.25,1.10);
  \draw[->, thick] (-1.13,0) -- (1.13,0) node[right] {\small $\mathrm{Re}\,\omega$};
  \draw[->, thick] (0,-1.2) -- (0,1.05) node[above] {\small $\mathrm{Im}\,\omega$};
  \draw[thick] (0,0) -- (0,-1);
  \draw[thick] (-1,0) -- (1,0);
  \definecolor{mygray}{RGB}{120,120,120}
  \draw[dotted, thick, mygray] (-1.05,0.05) arc[start angle=180, end angle=270, radius=1];
  \draw[dotted, thick, mygray] (1.05,0.05) arc[start angle=0, end angle=-90, radius=1];
  \draw[thick, mygray] (0.07,0.0) -- (0.07,-0.95);
  \draw[thick, mygray] (-0.07,-0.0) -- (-0.07,-0.95);
  \draw[thick, mygray] (-1.05,0.05) -- (-0.01,0.05);
  \draw[thick, mygray] (0.01,0.05) -- (1.05,0.05);
  \draw[thick, mygray] (0.02,0.05) arc[start angle=90, end angle=0, radius=0.05];
  \draw[thick, mygray] (-0.02,0.05) arc[start angle=90, end angle=180, radius=0.05];
  \foreach \n in {1,2}{
    \begin{scope}[shift={(0.55/\n,0.30*\n)}]
      \draw[very thick] (-0.026,-0.026) -- (0.026,0.026);
      \draw[very thick] (-0.026,0.026) -- (0.026,-0.026);
    \end{scope}
    \begin{scope}[shift={(-0.55/\n,0.30*\n)}]
      \draw[very thick] (-0.026,-0.026) -- (0.026,0.026);
      \draw[very thick] (-0.026,0.026) -- (0.026,-0.026);
    \end{scope}
  }
\end{tikzpicture}
}}}
\end{tabular}
\caption{Contour deformations of the filtered Green's function in the three time regimes: $u<-r_*$ (a), $-r_*<u<|r_*|$ (b), and $u>|r_*|$ (c). Crosses in the upper half plane denote the filter poles $\{\bar\omega_n,-\omega_n\}$. Solid gray contours indicate nonvanishing contributions, while dotted gray contours indicate vanishing ones. Panel (a) corresponds to the anti-causal contribution, which gives rise to the direct wave.}
\label{fig:causal_decomposition}
\end{figure*}

To see how the filter and source combine to produce $\Psi^{\rm filt}_{\ell m}$, we examine the causal structure of $G^{\rm filt}_\ell(u;r)$ by deforming the frequency integral in Eq.~\eqref{eq:G_filt_def} into the complex $\omega$ plane. For a source at $r_*$ and $u>|r_*|$, the contour closes in the lower half plane (LHP) \cite{Leaver:1986gd}, as illustrated in Fig.~\ref{fig:unfiltered_contour}.
The unfiltered Green's function has poles at the zeros of $A^{\rm inc}_{\rm in}$, i.e., the QNMs. The filter's zeros at $\{\omega_n,-\bar{\omega}_n\}$ cancel the selected QNM poles, while its poles at $\{\bar{\omega}_n,-\omega_n\}$ introduce mirror singularities in the upper half plane (UHP; crosses in the figure). Lying outside the LHP contour, these new poles do not contribute: $\Psi^{\rm filt}_{\ell m}$ is sourced solely by the branch cut along the negative imaginary axis (solid gray curve)\footnote{The large arc vanishes as indicated by the dotted circles.}.

At intermediate times $-r_*<u<|r_*|$, we follow \cite{Arnaudo:2025uos,Su:2026fvj,Kuntz:2025gdq} and split
\begin{align}
    G^{\rm filt}_{\ell}=G^{\rm filt,-}_\ell+G^{\rm filt,+}_\ell, \label{eq:G_filt_pm}
\end{align}
with, in the frequency domain,
\begin{align}
     & G^{\rm filt,-}_\ell=\frac{F(\omega)R_{\omega}^{\rm down}(r)}{2i\omega }, & G^{\rm filt,+}_{\ell}=\frac{F(\omega) A_{\rm in}^{\rm ref} R_{\omega}^{\rm up}(r)}{2i\omega A^{\rm inc}_{\rm in}}.
\end{align}
Here $R^{\rm up}_{\omega}\sim e^{i\omega r_*}$ and $R^{\rm down}_{\omega}\sim e^{-i\omega r_*}$ ($r_*\to+\infty$) are homogeneous solutions, with $R^{\rm in}_{\omega}=A^{\rm inc}_{\rm in}R^{\rm down}_{\omega}+A^{\rm ref}_{\rm in}R^{\rm up}_{\omega}$.
The two terms admit independent contour deformations, as illustrated in Fig.~\ref{fig:filtered_contours}.
For $G^{\rm filt,-}_\ell$, the contributions from the negative imaginary axis and the large arc both vanish~\cite{Su:2026fvj} (dotted curves), so closing in the LHP leaves only the small arc around $\omega=0$ (solid curve). For $G^{\rm filt,+}_\ell$, the contour instead closes in the UHP, picking up the positive-imaginary-axis branch cut together with the poles of $F(\omega)$.
Here, the pole contribution is
\begin{align}
    &2{\rm Re}\,\sum_{n=0}^N E_n A_{\rm in}^{\rm ref} (\bar{\omega}_n) R_{\bar{\omega}_n}^{\rm up}(r)e^{-i\bar{\omega}_n u},
    & -r_*<u<|r_*|,
\end{align}
with excitation factors
\begin{align}
    E_n=\lim_{\omega\to\bar\omega_n}(\omega-\bar\omega_n)\,\frac{F(\omega) }{2\omega A^{\rm inc}_{\rm in}}.
\end{align}

The decomposition in Eq.~\eqref{eq:G_filt_pm} avoids the high-frequency large arc, reducing the calculation to numerically tractable pieces --- the small arc, the branch cut, and the filter poles. This strategy recently enabled a first-principles computation of the prompt response in Schwarzschild spacetime~\cite{Ma:2026prompt}.

At early times $u<-r_*$, no split is needed: closing the contour in the UHP (Fig.~\ref{fig:uhp_no_pia}), the filtered Green's function is determined entirely by the UHP poles\footnote{The Green's function has no branch cut along the positive imaginary axis, and the {large-arc contribution} (dotted circle) vanishes.}:
\begin{align}
    &G^{\rm filt}_{\ell,{\rm anti-causal}}(u;r)=2{\rm Re}\,\sum_{n=0}^N E_n R_{\bar{\omega}_n}^{\rm in}(r)e^{-i\bar{\omega}_n u}, \quad  u<-r_*. \label{eq:Green_function_anti_poles}
\end{align}
By causality, the unfiltered retarded Green's function vanishes in this window; Eq.~\eqref{eq:Green_function_anti_poles} is thus anti-causal: the filter endows the retarded Green's function with an advanced component propagating outside the original light cone.

Collecting the three windows, the total filtered Green's function reads
\begin{align}
    G^{\rm filt}_\ell(u;r)&=G^{\rm filt}_{\ell,\,{\rm anti-causal}}(u;r)\, \Theta(-u-r_*) \notag \\
    &+G^{\rm filt}_{\ell,\,{\rm prompt}}(u;r) \,\Theta(u+r_*) \,\Theta(-u+|r_*|) \notag \\
    &+G^{\rm filt}_{\ell,\,{\rm tail}}(u;r) \, \Theta(u-|r_*|). \label{eq:filtered_green_function_decomposition}
\end{align}
This decomposition provides a first-principles framework for how filtered waveforms, direct waves in particular, are assembled from anti-causal, prompt, and tail components.

\section{Decomposing Zerilli waveforms and modeling direct waves}
Inserting Eq.~\eqref{eq:filtered_green_function_decomposition} into Eq.~\eqref{eq:Psi_Zerilli_convolution}, the filtered Zerilli waveform separates into
\begin{align}
    \Psi^{{\rm filt}}_{\ell m}(u)&=\mathop{\iint}\limits_{\substack{t-r_*>u}} dt\, dr\,  G^{\rm filt}_{\ell,\,{\rm anti-causal}}\bigl(u - t; r\bigr) S_{\ell m}(t,r) \notag \\
    &+\mathop{\iint}\limits_{\substack{t-r_*<u<t+|r_*|}} dt\, dr\,  G^{\rm filt}_{\ell,\,{\rm prompt}}\bigl(u - t; r\bigr) S_{\ell m}(t,r) \notag \\
    &+\mathop{\iint}\limits_{\substack{t+|r_*|<u}} dt\, dr\,  G^{\rm filt}_{\ell,\,{\rm tail}}\bigl(u - t; r\bigr) S_{\ell m}(t,r). \label{eq:filtered_zerilli_decomposition}
\end{align}
We now specialize to a point-particle source, which takes the form~\cite{Nagar:2006xv}
\begin{align}
    S_{\ell m}(t,r)=A(r,t)\delta(r-R(t))+B(r,t)\partial_{r}\delta(r-R(t)), \label{eq:Zerilli_source_formal}
\end{align}
where $R(t)$ denotes the particle trajectory.

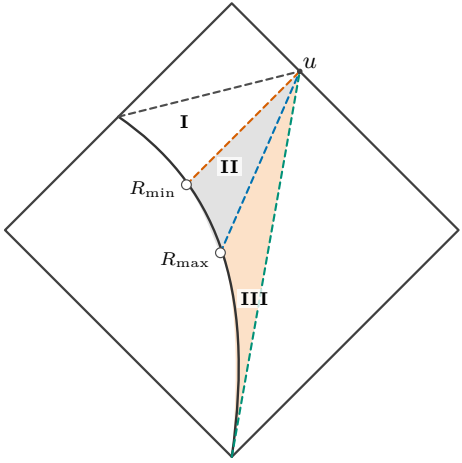
\begin{figure}[!htb]
\begin{tikzpicture}[
  scale=3,
  line cap=round,
  line join=round,
  boundary/.style={draw=black!75, line width=0.85pt},
  horizon/.style={draw=black!80, line width=0.9pt},
  guide/.style={line width=0.8pt, dash pattern=on 2.4pt off 1.7pt},
  marker/.style={circle, fill=white, draw=black!70, line width=0.45pt, inner sep=1.3pt},
  regionlabel/.style={fill=white, fill opacity=0.78, text opacity=1, inner sep=1.1pt}
]
\definecolor{promptshade}{RGB}{226,226,226}
\definecolor{tailshade}{RGB}{252,225,200}
\definecolor{blueguide}{RGB}{0,114,178}
\definecolor{greenguide}{RGB}{0,145,120}
\definecolor{vermillionguide}{RGB}{213,94,0}

\coordinate (N) at (0,1);
\coordinate (E) at (1,0);
\coordinate (S) at (0,-1);
\coordinate (W) at (-1,0);
\coordinate (A) at (-0.2,0.2);
\coordinate (B) at (0.3,0.7);
\coordinate (C) at (-0.05,-0.1);
\coordinate (D) at (-0.5,0.5);

\fill[promptshade] (A) -- (B) -- (C)
  .. controls (-0.08,-0.05) and (-0.12,0.1) .. (A) -- cycle;
\fill[tailshade] (B) -- (S)
  .. controls (0.05,-0.4) and (-0.02,-0.15) .. (C)
  -- cycle;

\draw[boundary] (W) -- (N) -- (E) -- (S) -- cycle;
\draw[horizon] (D)
  .. controls (-0.08,0.21) and (0.11,-0.23) .. (S);

\draw[guide, black!70] (D) -- (B);
\draw[guide, vermillionguide] (A) -- (B);
\draw[guide, blueguide] (C) -- (B);
\draw[guide, greenguide] (S) -- (B);

\node[regionlabel] at (-0.21,0.48) {\scriptsize\bfseries I};
\node[regionlabel] at (-0.01,0.28) {\scriptsize\bfseries II};
\node[regionlabel] at (0.10,-0.30) {\scriptsize\bfseries III};

\fill[black!75] (B) circle (0.012);
\node[above right, inner sep=1pt] at (B) {$u$};
\node[marker] at (A) {};
\node[anchor=east, inner sep=0.6pt] at (-0.235,0.18) {\scriptsize $R_{\rm min}$};
\node[marker] at (C) {};
\node[anchor=east, inner sep=0.6pt] at (-0.085,-0.13) {\scriptsize $R_{\rm max}$};

\end{tikzpicture}
    \caption{Penrose diagram illustrating the causal decomposition of a filtered waveform. A particle plunges into the future horizon along the black trajectory. The filtered signal observed at future null infinity at retarded time $u$ receives contributions from three trajectory segments, separated by $R_{\rm min}$ and $R_{\rm max}$. Segment~I is anti-causal and sources the direct wave, while segments~II and III give the prompt response and tail, respectively. The red dashed line denotes the outgoing light cone.}
    \label{fig:waveform_decomposition}
\end{figure}

Figure~\ref{fig:waveform_decomposition} illustrates this decomposition: for a particle falling into the future horizon, the filtered waveform observed at $u$ receives contributions from three trajectory segments, separated by $R_{{\rm min}}$ and $R_{{\rm max}}$ satisfying
\begin{align}
    &t(R_{{\rm min}})-R_{*,{\rm min}}=u,  \notag \\
    &t(R_{{\rm max}})+|R_{*,{\rm max}}|=u, \notag
\end{align}
where $t(R)$ is the inverse of $R(t)$, and the red dashed line connecting $R_{\rm min}$ to $u$ marks the outgoing light cone. {The anti-causal contribution, sourced by segment~I near the future horizon, reads}
\begin{align}
    2{\rm Re}\,\sum_{n=0}^N &E_n e^{-i\bar{\omega}_n u} \int^{R_{\rm min}(u)}_{2} \frac{dR}{|\dot{R}|} e^{i\bar{\omega}_n t(R)}\notag \\
    &\times [R_{\bar{\omega}_n}^{\rm in}(A-\partial_rB)-B\partial_r R_{\bar{\omega}_n}^{\rm in}]_{t=t(R),r=R}, \label{eq:anti_causal_source_convolution}
\end{align}
where we have inserted Eqs.~\eqref{eq:Green_function_anti_poles} and \eqref{eq:Zerilli_source_formal} into Eq.~\eqref{eq:filtered_zerilli_decomposition}.

\begin{figure}[!htb]
    \includegraphics[width=\columnwidth]{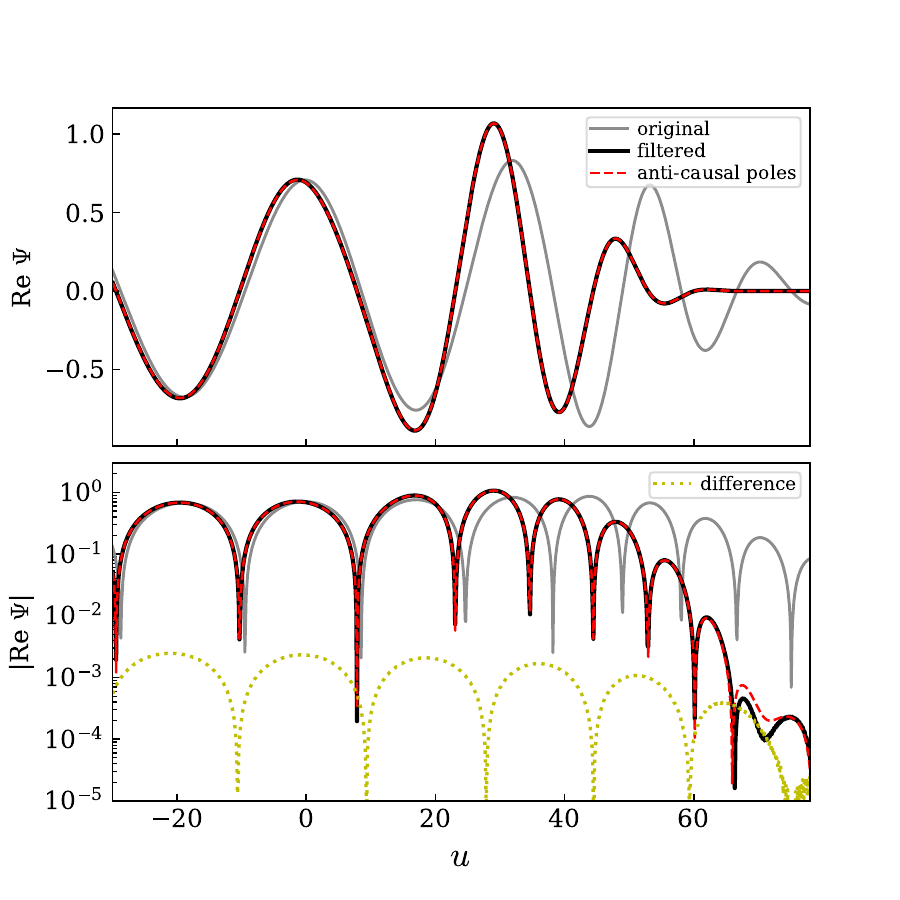}
    \caption{The $\ell=m=2$ Zerilli waveform for an ISCO plunge: unfiltered (gray) and filtered with overtones up to $n=5$ removed (black). The decaying cycles over $u\in[40,60]$ correspond to direct waves. The anti-causal contribution alone [red dashed, Eq.~\eqref{eq:anti_causal_source_convolution}] reproduces the filtered waveform, up to a $\sim0.1\%$ residual (yellow dashed). 
    }
    \label{fig:fig_zerilli_assembly_shifted_real}
\end{figure}

In Fig.~\ref{fig:fig_zerilli_assembly_shifted_real}, we consider a particle plunging from the innermost stable circular orbit (ISCO), with specific energy $2\sqrt{2}/3$ and angular momentum $2\sqrt{3}$. The gray and black curves show the $\ell=m=2$ harmonic of the unfiltered and filtered Zerilli waveforms, respectively. The rational filter, here including overtones up to  $n=5$, efficiently removes the dominant QNM oscillation.
Over $u\in[40,60]$, the filtered waveform displays several decaying cycles, identified as ``direct waves'' in \cite{Oshita:2025qmn}.

Decomposing the filtered waveform via Eq.~\eqref{eq:filtered_zerilli_decomposition}, we find, remarkably, that the anti-causal piece alone [red dashed, Eq.~\eqref{eq:anti_causal_source_convolution}] closely reproduces the full filtered waveform; the prompt response and tail together contribute only $\sim0.1\%$ (yellow dashed). We have verified that this conclusion remains valid for other orbital configurations, including eccentric plunges.

Each anti-causal pole $\bar{\omega}_n$ in Eq.~\eqref{eq:anti_causal_source_convolution} contributes much like the dynamical excitation of QNMs~\cite{DeAmicis:2025xuh}, which takes the form
\begin{align}
    \sim \sum_{n=0}^N & e^{-i{\omega}_n u} \int_{R_{\rm max}(u)}^{+\infty} \frac{dR}{|\dot{R}|} e^{i{\omega}_n t(R)}\notag \\
    &\times [R_{{\omega}_n}^{\rm in}(A-\partial_rB)-B\partial_r R_{{\omega}_n}^{\rm in}]_{t=t(R),r=R}. \label{eq:dynamical_qnm} 
\end{align}
Comparing Eqs.~\eqref{eq:dynamical_qnm} and \eqref{eq:anti_causal_source_convolution} is instructive. The integration domain of Eq.~\eqref{eq:dynamical_qnm} shows that the dynamical QNM excitation samples the source in the chronological past of $u$ (Fig.~\ref{fig:waveform_decomposition}). As $u$ increases, Eq.~\eqref{eq:dynamical_qnm} asymptotes to the usual constant-amplitude QNM contribution, $\sum_n {\rm (const)} \times e^{-i\omega_n u}$, plus a dynamical piece.
The filtered waveform, by contrast, draws its trajectory information from the complementary anti-causal segment (region~I in Fig.~\ref{fig:waveform_decomposition}). The oscillation and decay of the direct wave are therefore controlled by the near-horizon dynamics of the source.

Near the horizon, the second line of Eq.~\eqref{eq:anti_causal_source_convolution} scales as $(R-2)e^{-i\bar{\omega}_nR_*}$, as follows from Eq.~\eqref{eq:in_solution} and the near-horizon behavior of the source terms $A$ and $B$. Together with $|\dot R|\sim R-2$, the leading contribution in Eq.~\eqref{eq:anti_causal_source_convolution} is
\begin{align}
    \sim \sum_n e^{-i\bar{\omega}_n u} \int^{R_{\rm min}(u)}_{2} dR e^{i\bar{\omega}_n t(R)} e^{-i\bar{\omega}_n R_*} e^{-im\Phi}, \label{eq:near_horizon_expansion}
\end{align}
where $\Phi$ is the orbital phase of the source.
{Integrating by parts and dropping higher-order corrections}\footnote{A similar treatment was used to study dynamical tides of neutron stars; see Appendix A of \cite{Ma:2020rak}.}, this reduces to
\begin{align}
    \sim e^{-i\int \omega_G\, du}, \label{eq:near_horizon_expansion_ibp}
\end{align}
with 
\begin{align}
    \omega_G=m\frac{d\Phi/dt}{1+|dR_*/dt|}-2i\kappa\frac{|dR_*/dt|}{1+|dR_*/dt|}.
\end{align}
This $\omega_G$ coincides with the quantity introduced by Oshita \etal \cite{Oshita:2025qmn} [Eq.~(5) therein], later used to analyze the direct-wave component in GW250114~\cite{Lu:2025vol}.

However, relating the present derivation to \cite{Oshita:2025qmn} requires care. There, the integral in Eq.~\eqref{eq:Psi_Zerilli_convolution} was argued to decompose into QNMs and direct waves, with the latter estimated by steepest descent. That argument is best viewed as heuristic --- it did not specify the contour deformation in the complex-frequency plane, and it led to the incorrect interpretation of the direct wave as a light-cone contribution propagating along the red dashed line in Fig.~\ref{fig:waveform_decomposition}.

The first-principles expression in Eq.~\eqref{eq:anti_causal_source_convolution} instead shows that the direct wave is an accumulated effect from the anti-causal segment extending down to the horizon; it is the boundary term from the integration by parts in Eq.~\eqref{eq:near_horizon_expansion_ibp} that makes it appear instantaneous.
It was recently argued \cite{Kuntz:2026xep} that the light-cone (instantaneous) wave may vanish in Schwarzschild spacetime. Our analysis clarifies why that cancellation does not apply to the direct wave in \cite{Oshita:2025qmn}.

\begin{figure}[!htb]
    \includegraphics[width=\columnwidth]{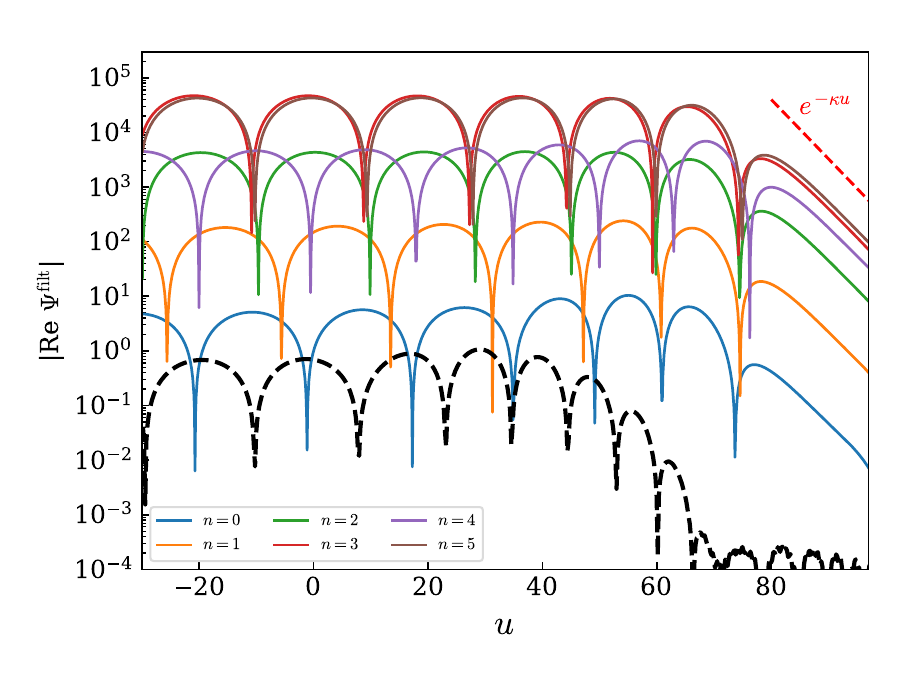}
    \caption{Individual anti-causal poles $\bar{\omega}_n$ contributing to the direct wave [solid curves, Eq.~\eqref{eq:anti_causal_source_convolution}]. Each pole decays at late times as a horizon mode $\sim e^{-\kappa u}$, shown by the red dashed line for reference. They cancel collectively in the total filtered waveform (black dashed).}
    \label{fig:fig_zerilli_early_components}
\end{figure}

Equation~\eqref{eq:anti_causal_source_convolution} expresses the direct wave as a superposition of anti-causal poles $\bar{\omega}_n$, whose individual time evolution is shown as solid curves in Fig.~\ref{fig:fig_zerilli_early_components}. {At late times} each contribution decays as a horizon mode $\sim e^{-\kappa u}$ \cite{Mino:2008at,Zimmerman:2011dx}, with $\kappa=1/4$ the surface gravity --- behavior also seen in the dynamical aspect of QNMs \cite{DeAmicis:2025xuh}. However, these horizon modes cancel collectively in the total filtered waveform (black dashed curve): the direct-wave pattern emerges only from the collective sum of the $\bar{\omega}_n$ contributions.

\begin{figure}[!htb]
    \includegraphics[width=\columnwidth]{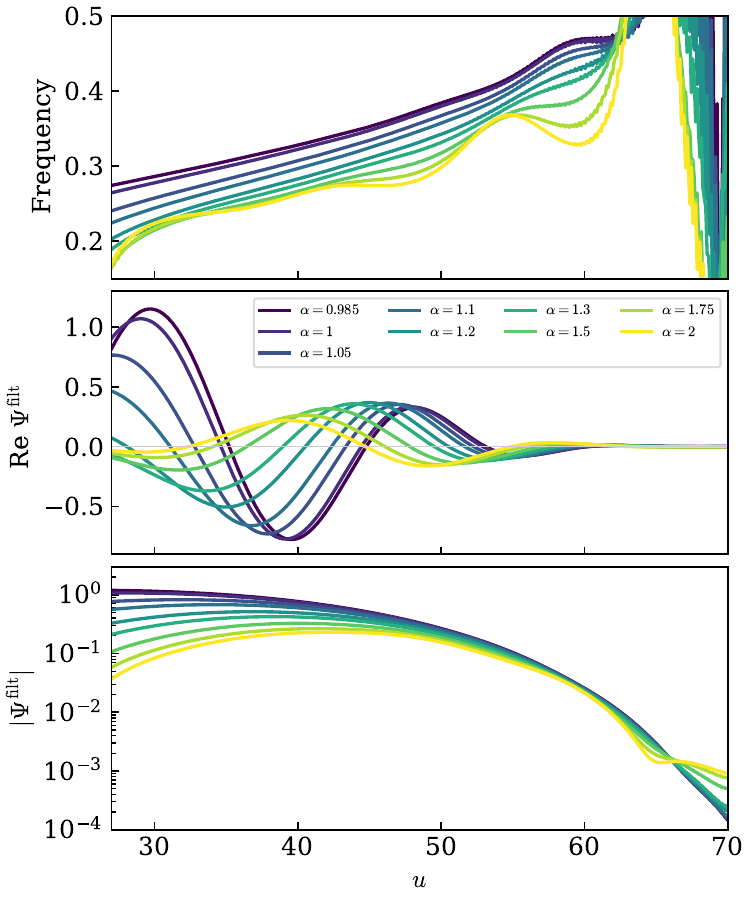}
    \caption{Dependence of the direct wave on the near-horizon source motion under the phase deformation $\Phi(R)\to\Phi(R/\alpha)$. Shown are the instantaneous frequency (top), real part of the filtered waveform (middle), and amplitude (bottom). Increasing $\alpha$ slows the orbital phase evolution and systematically lowers the instantaneous frequency of the direct waves.}
    \label{fig:fig_phasefac_omega_re_abs_stack}
\end{figure}

Equation~\eqref{eq:anti_causal_source_convolution} also explicitly connects the direct-wave signal to the near-horizon source. To demonstrate this, we deform the source trajectory by rescaling the orbital phase, $\Phi(R)\to \Phi(R/\alpha)$, where the constant $\alpha$ slows the phase evolution.
As shown in Fig.~\ref{fig:fig_phasefac_omega_re_abs_stack}, increasing $\alpha$ systematically lowers the instantaneous frequency of the direct-wave cycles and shifts their phase, supporting that near-horizon source dynamics are correlated with direct waves.

\section{Discussion}
\label{sec:conclusion}
By analyzing the causal structure of the filtered Zerilli Green's function, we have laid a first-principles foundation for the direct waves identified in \cite{Oshita:2025qmn,Lu:2025vol}. The direct wave is sourced by the anti-causal trajectory segment extending down to the future horizon, making it an accumulated near-horizon effect, not a light-cone (instantaneous) contribution. It is therefore not subject to the cancellation mechanism of \cite{Kuntz:2026xep}. 

Although the direct wave arises from the poles of the filter, whose profile is absent from the unfiltered time-domain waveform, it is not a spurious artifact. The filter has unit modulus, $|F(\omega)|=1$ (for real $\omega$), so the source information encoded in
\begin{align}
    \Psi(\omega)=\int G_\omega(r)S_{\omega}(r) dr
\end{align}
is preserved: $F(\omega)$, being source independent, merely redistributes it in time by reshaping the causal structure of the Green's function. The direct-wave cycles are thus not new physical content, but a time-domain rearrangement of information already in $\Psi(\omega)$.

Our analysis further suggests that, at least for Schwarzschild, the horizon modes cancel collectively in the total filtered waveform. The direct wave should therefore not be modeled as a horizon mode, as in Ref.~\cite{Chung:2026eph}; its instantaneous frequency and decay rate are instead fixed by the near-horizon source dynamics, controlled by the plunge initial conditions and gradually modulated by near-horizon redshift and, in rotating spacetimes, frame dragging. Thus, while direct waves do not generally coincide with horizon modes~\cite{Dyer:2026yex,Kankani:2026byb,Kankani:2026kst}, they still encode how horizon signatures gradually enter the source dynamics \cite{Oshita:2025qmn}. Constructing the map between horizon signatures and direct-wave observables is an important next step.

Although restricted to linear perturbations of Schwarzschild, our analysis is a concrete step toward understanding direct waves in filtered NR waveforms~\cite{Lu:2025vol,Oshita:2025qmn}, where similar decaying cycles consistently appear. In~\cite{Lu:2025vol}, {the direct waves of comparable-mass binaries were attributed, via an effective-one-body picture, to a source plunging into an effective BH spacetime}. Our results suggest a more fundamental interpretation.

Once a common horizon has formed, the geometry on a suitable Cauchy slice and in its future domain of dependence can be viewed as a perturbed remnant BH --- a picture supported by the successes of the close-limit approximation~\cite{Price:1994pm}, the Lazarus project~\cite{Baker:2001sf}, and the hybrid approach~\cite{Nichols:2010qi,Nichols:2011ih}, which treat suitable post-merger regions perturbatively while retaining agreement with NR. The inclusion of second-order effects can further improve this agreement \cite{Gleiser:1995gx,Gleiser:1996yc}.

In this picture, the residual post-merger distortion on the chosen Cauchy slice acts, after a Laplace transform, as an effective source in the perturbation equations, generating direct waves around the actual remnant BH, rather than an effective one. Sourced by the anti-causal segment, these cycles are governed by the later post-merger dynamics, where nonlinearities have largely decayed, and are hence amenable to BH perturbation theory. Extending our calculation to Kerr spacetime is therefore an essential next step for understanding direct waves in NR waveforms.

\begin{acknowledgments}
The calculations in this work were performed with the assistance of the autonomous scientific workflow \Chandra \cite{wang2026directed} based on the large language model \texttt{Claude Opus 4.7}.
We thank Yanbei Chen and Ling Sun for useful comments on the manuscript.
Research at Perimeter Institute is supported in part by the Government of Canada through the Department of Innovation, Science and Economic Development and by the Province of Ontario through the Ministry of Colleges, Universities, Research Excellence and Security.
This work used Delta at the
National Center for Supercomputing Applications (NCSA) through allocation CIS250856 from the Advanced Cyberinfrastructure Coordination Ecosystem: Services \& Support (ACCESS) program~\cite{10.1145/3569951.3597559}, which is supported by U.S. National Science Foundation grants $\#$2138259, $\#$2138286, $\#$2138307, $\#$2137603, and $\#$2138296.

\end{acknowledgments}  



\def\bibsection{\section*{References}}
\bibliography{References}

\end{document}